\def\year{2021}\relax
\documentclass[letterpaper]{article} 
\usepackage{aaai21}  
\usepackage{times}  
\usepackage{helvet} 
\usepackage{courier}  
\usepackage[hyphens]{url}  
\usepackage{graphicx} 
\usepackage{natbib}  
\usepackage{caption} 
\usepackage{booktabs} 

\usepackage[english]{babel}
\usepackage{moresize}
\usepackage{amsmath}

\usepackage{algorithmic} 
\usepackage{balance}
\usepackage{comment}
\usepackage{paralist}
\usepackage{multirow}

\usepackage{filecontents}

\usepackage[english]{babel}
\usepackage[latin1]{inputenc}
\usepackage{mathrsfs}
\usepackage{graphicx}
\usepackage{amssymb}
\usepackage{url}
\usepackage{subfigure}
\usepackage{amsmath}
\usepackage{enumitem}
\usepackage[linesnumbered,algoruled,boxed,lined]{algorithm2e}
\usepackage{adjustbox}
\usepackage{amssymb}
\usepackage{times}
\usepackage{hyperref}

\usepackage{pgfplots}
\usetikzlibrary{pgfplots.dateplot}

\usepackage{filecontents}
\definecolor{tblue}{RGB}{31,119,180}
\definecolor{torange}{RGB}{255,127,14}
\definecolor{tgreen}{RGB}{44,160,44}
\definecolor{tred}{RGB}{214,39,40}
\definecolor{tpurple}{RGB}{148,103,189}

\usepackage{filecontents}

\newcommand{\hide}[1]{} 

\newcommand{\ie}{\textit{i}.\textit{e}.}
\newcommand{\eg}{\textit{e}.\textit{g}.} 
\newcommand{\wrt}{\textit{w}.\textit{r}.\textit{t}} 
\newtheorem{Dfn}{Definition}

\title{Knowledge-aware Coupled Graph Neural Network for Social Recommendation}
\author{
    Chao Huang$^1$\thanks{Both authors contribute equally to this work}, Huance Xu$^{2*}$, Yong Xu$^{2,3,4}$\thanks{Corresponding author: Yong Xu}, Peng Dai$^1$, Lianghao Xia$^2$, \\\Large{\bf Mengyin Lu $^1$, Liefeng Bo$^1$, Hao Xing$^5$, Xiaoping Lai$^5$, Yanfang Ye$^6$}
    \\
}
\affiliations{
    \textsuperscript{\rm 1}JD Finance America Corporation, USA\\
    \textsuperscript{\rm 2}South China University of Technology, China, \textsuperscript{\rm 3}Peng Cheng Laboratory, China\\
    \textsuperscript{\rm 4}Communication and Computer Network Laboratory of Guangdong, China\\
    \textsuperscript{\rm 5}VIPS Research, China, \textsuperscript{\rm 6}Case Western Reserve University, USA\\
    chaohuang75@gmail.com, \{cshuance.xu, cslianghao.xia\}@mail.scut.edu.cn, yxu@scut.edu.cn, \\\{peng.dai,mengyin.lu,liefeng.bo\},
\{hao.xing,tom.lai\}vipshop.com, yanfang.ye@case.edu

}

\def\model{KCGN}

\begin{document}

\maketitle

\begin{abstract}
Social recommendation task aims to predict users' preferences over items with the incorporation of social connections among users, so as to alleviate the sparse issue of collaborative filtering. While many recent efforts show the effectiveness of neural network-based social recommender systems, several important challenges have not been well addressed yet: (i) The majority of models only consider users' social connections, while ignoring the inter-dependent knowledge across items; (ii) Most of existing solutions are designed for singular type of user-item interactions, making them infeasible to capture the interaction heterogeneity; (iii) The dynamic nature of user-item interactions has been less explored in many social-aware recommendation techniques. To tackle the above challenges, this work proposes a \underline{\textbf{K}}nowledge-aware \underline{\textbf{C}}oupled \underline{\textbf{G}}raph \underline{\textbf{N}}eural Network (\model) that jointly injects the inter-dependent knowledge across items and users into the recommendation framework. \model\ enables the high-order user- and item-wise relation encoding by exploiting the mutual information for global graph structure awareness. Additionally, we further augment \model\ with the capability of capturing dynamic multi-typed user-item interactive patterns. Experimental studies on real-world datasets show the effectiveness of our method against many strong baselines in a variety of settings. Source codes are available at: \emph{https://github.com/xhcdream/KCGN}.
\end{abstract}

\section{Introduction}
\label{sec:intro}

In recent years, social recommendation which aims to exploit users' social information for modeling users' preferences in recommendations, has attracted significant attention~\cite{liu2019discrete}. As has been stated in many social-aware recommendation literature~\cite{wu2019neural,chen2019efficient}, social influences between users have high impacts on users' interactive behavior over items in various recommender scenarios, such as e-commence~\cite{lin2019cross} and online review platforms~\cite{chen2020social}. Hence, researchers propose to incorporate social ties into the collaborative filtering architecture as side information to characterize connectivity information across users.

The most common paradigm for state-of-the-art social recommender systems is to learn an embedding function, which unifies user-user and user-item relations into latent representations. To tackle this problem, many studies have developed various neural network techniques to integrate social information with the user-item interaction encoding as constraints. For example, attention-based mechanism has been utilized to aggregate correlations among different users~\cite{chen2019social,chen2019efficient}. Furthermore, inspired by the recent advance of graph neural architectures, several attempts are built upon the message passing frameworks over the user-user social graph. For example, social influence is simulated with layer-wise diffusion scheme for information fusion~\cite{wu2019neural}. GraphRec~\cite{fan2019graph} employs the graph attention network to model the relational structures between users. To enable the modeling context-aware social effects, DANSER~\cite{wu2019dual} stacks two-stage of graph attention layer for distinguishing the multi-faceted social homophily and influence.

While these solutions have provided encouraging results, several key aspects have not been well addressed yet. In particular, \emph{First}, in real-life scenarios, there typically exist relations between items which characterize item-wise fruitful semantics relatedness, and are helpful to understand user-item interactive patterns~\cite{wang2019knowledge}. For instance, in online retailing systems, products of the same categories (\eg, food \& grocery, clothing \& shoes) or complement with each other, could be correlated to enrich the knowledge representation of items~\cite{xin2019relational}. For online review platforms, the exploiting of dependencies among the venues with the same functionality, is able to provide external knowledge in assisting user preference learning~\cite{yu2019multi}. However, the majority of existing social recommender systems fail to capture item-wise relational structures, which can hardly distill the knowledge-aware collaborative signals from the co-interactive behaviors of users.

\emph{Second}, to simplify the model design, most of current social recommendation methods have thus far focused on modeling singular type of interactive relations between user and item. Yet, many practical recommendation scenarios may involve the diversity of users' interaction over items~\cite{cen2019representation,xia2020multiplex}. Take the e-commerce site as an example, the effective encoding of multi-typed user-item interactive patterns (\eg, page view, add-to-favorite and purchase) and their underlying inter-dependencies (\eg, add-to-favorite activities may serve as useful indicators for making purchase decisions), is crucial to more accurately inference of user's complex interest in social recommendation tasks.

\emph{Third}, the time dimension of the social recommendation deserves more investigation, so as to capture behavior dynamics. Most of recent approaches ignore the dynamic nature of user-item interactions and assume that the factor influencing the interactive behavior is only the identity of items~\cite{song2019session}. While there exist a handful of recent work that consider the sequential information in social recommendation~\cite{song2019session,sun2018attentive}, their are limited in their intrinsic design for singular type of user-item relations. This makes them insufficient to yield satisfactory embeddings with the preservation of different interaction signals in a dynamic manner for more complex scenarios.

While intuitively useful to integrate the above dimensions into social recommendation frameworks, two unique technical challenges arise in achieving this goal. Specifically, graph-structured neural network can be applied to naturally model the topological information of social node instances, such as the graph-based convolutional network~\cite{wu2019neural} or attention mechanism~\cite{wu2019dual,fan2019graph}. However, their non-linear aggregation functions can only learn the local proximity between users and are incapable of capturing the broader context of the graph structure (\eg, users with the isomorphic social structures)~\cite{you2019position}. Hence, how to jointly capture knowledge-aware user-user and item-item local relations, as well as retain the high-order social influence and item dependencies under global context, remains a significant challenge. Additionally, it is also very challenging to handle the dynamic multi-typed user-item interactions, so as to capture the dynamic relation-aware structural dependencies across users and items with arbitrary duration.\\\vspace{-0.12in}

\noindent \textbf{The Present Work}. In light of the aforementioned motivations and challenges, we study the social recommendation problem by proposing the \underline{\textbf{K}}nowledge-aware \underline{\textbf{C}}oupled \underline{\textbf{G}}raph \underline{\textbf{N}}eural Network (\model). To jointly deal with the user-user and item-item local and global relational structure awareness, we incorporate the mutual information estimation schema into the coupled graph neural architecture. This design enables the collaboration between neural mutual information estimator and graph-structured representation learning paradigm, which preserves the node-level unique characteristics and graph-level substructure knowledge across users and items. In addition, to capture the dynamic multi-typed interactive patterns, we integrate a relation-aware message passing framework with the relative temporal encoding strategy, which endows \model\ with the capability of incorporating the temporal information into the multi-typed user-item interaction graph learning.



Our contributions can be highlighted as follows:
\begin{itemize}[leftmargin=*]

\item We propose to capture both user-user and item-item relations with the developed coupled graph neural network. Through the joint modeling of user- and item-wise dependent structures, our \model\ can enhance the social-aware user embeddings with the preservation of knowledge-aware cross-item relations in a more thorough way.

\item We propose a relation-aware graph neural module to encode the multi-typed user-item interactive patterns, and further incorporate the temporal information into the message passing kernel to enhance the learning of collaborative relations for recommendation.

\item We conduct extensive experiments on three real-world datasets to show the superiority of our \model\ when competing with several baselines from various research lines. Further studies on scalability evaluation validate the model efficiency of \model\ over state-of-the-art social recommender systems. We also show that our model maintains strong performance in the cold-start scenarios when user-item interactions are sparse.
\end{itemize}

\section{Problem Definition}
\label{sec:model}

We first introduce key definitions of social recommendation with item relational knowledge and different types of user-item interactions. We consider a typical recommendation scenario, in which we have $I$ users $U=\{u_1,...,u_i,...,u_I\}$ and $J$ items $V=\{v_1,...,v_j,...,v_J\}$. To capture the multi-typed user-item interaction signals, we define a multi-typed interaction tensor as below:

\begin{Dfn}
\textbf{Multi-typed Interaction Tensor} $\textbf{X}$. We define a three-way tensor $\textbf{X} \in \mathbb{R}^{I\times J\times K}$ to represent the different types of interactions between user and item, where $K$ (indexed by $k$) denotes the number of interaction types (page view, purchase, or like, dislike). In $\textbf{X}$, the element $x_{i,j}^k=1$ if user $u_i$ interacts with item $v_j$ with the interaction type of $k$ and $x_{i,j}^k=0$ otherwise. To deal with the interaction dynamics, we also define a temporal tensor $\textbf{T} \in \mathbb{R}^{I\times J\times K}$ with the same size of $\textbf{X}$ to record the timestamp information ($t_{i,j}^k$) of each corresponding interaction $x_{i,j}^k$.
\end{Dfn}

\begin{Dfn}
\textbf{User Social Graph} $G_u$. $G_u=\{U,E_u\}$ represents the social relationships (edges $E_u$) among users (nodes $U$), where there exists an edge $e_{i,i'}$ between user $u_i$ and $u_{i'}$ given they are socially connected.
\end{Dfn}

\begin{Dfn}
\textbf{Item Inter-Dependency Graph} $G_v$. We further define $G_v=\{V,E_v\}$ to represent the inter-dependence of items. In particular, we characterize the item-wise relations with a triple $\{v_j, e_{j,j'}, v_{j'} | v_j, v_{j'} \in V \}$, where edge $e_{j,j'}$ describes the relationship between item $v_j$ and $v_{j'}$, \eg, $v_j$ and $v_{j'}$ belong to the same product categories and have similar functionality, or are interacted by the same user under the same interaction type of $k$.
\end{Dfn}

\noindent \textbf{Task Formulation}. We formulate the studied recommendation task in this paper as: \textbf{Input}: multi-typed interaction tensor $\textbf{X} \in \mathbb{R}^{I\times J\times K}$, user social graph $G_u$ and item inter-dependence graph $G_v$. \textbf{Output}: a predictive function that effectively forecasts the future user-item interaction.

\section{Methodology}
\label{sec:solution}


\begin{figure}
	\centering
	\includegraphics[width=0.49\textwidth]{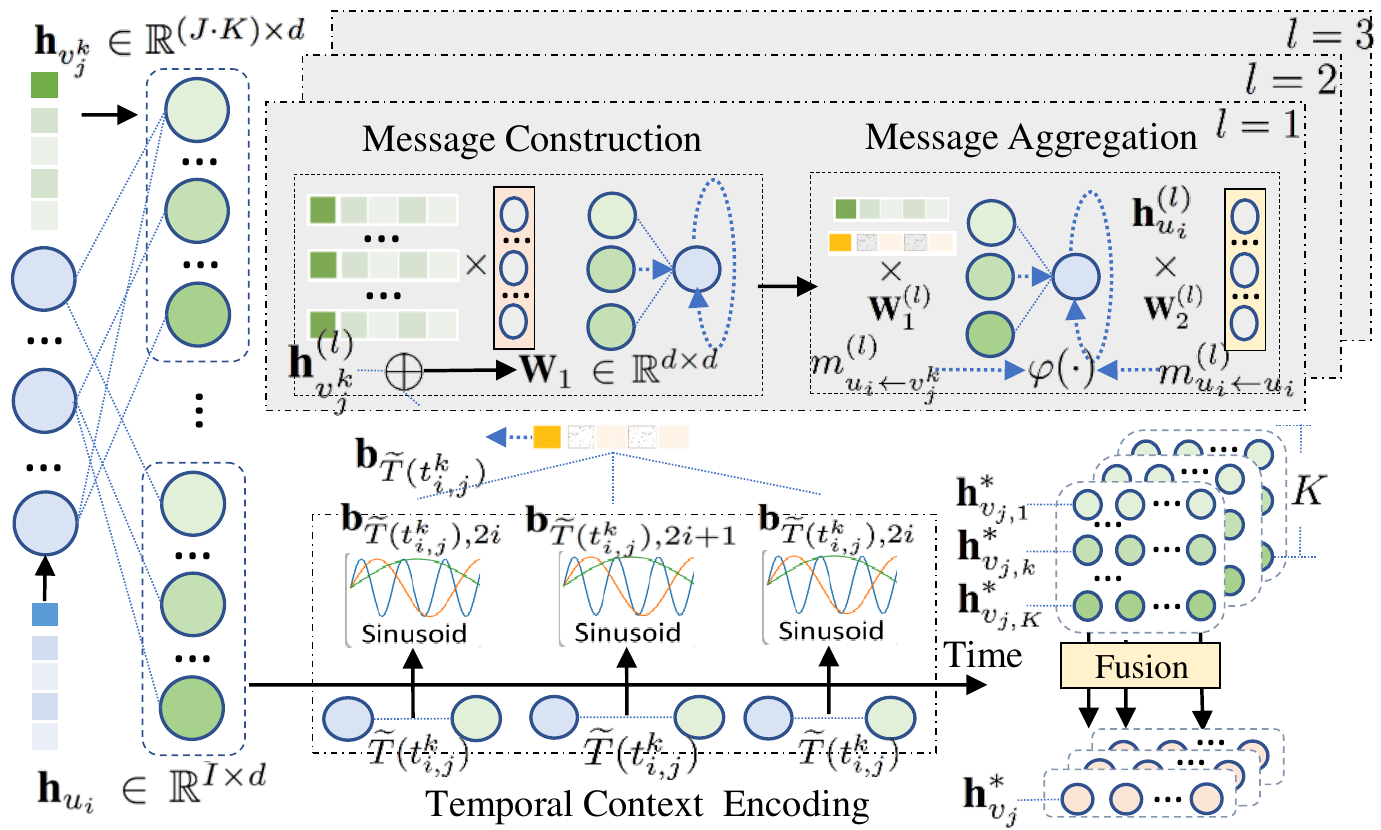}
	\caption{The architecture of the multi-typed interactive pattern modeling. $\oplus$ denotes the element-wise addition.}
	\label{fig:framework_1}
	\vspace{-0.15in}
\end{figure}

\subsection{Multi-typed Interactive Pattern Modeling}
To encode the multi-typed collaborative relations, we propose a relation-aware graph neural architecture, which is built upon the message passing paradigm (as shown in Figure~\ref{fig:framework_1}), to empower \model\ to capture the dedicated patterns of different types of user-item interactions. Specifically, given the multi-typed interaction tensor $\textbf{X}$, we first construct a multi-typed relation graph $G_m$ by representing the interaction heterogeneity with type-specific item sub-vertices $v_j \rightarrow (v_j^1,...,v_j^k,...,v_j^K)$, where $K$ denotes the number of interaction types. Each edge between $u_i$ and $v_j^k$ represents the corresponding interaction with the $k$-th type. After that, there are $(I+J\cdot K)$ vertices in our multi-typed graph $G_m=(V_m,E_m)$, where $V_m=U\cup V'$ and $v_j^k \in V'$. Here, $V'$ is the new type-aware item set.

\subsubsection{Message Construction Phase.}
We first generate the message between user vertex $u_i$ and his/her interacted type-specific item vertex $v_j^k$ as follows:
\begin{align}
\label{eq:message_passing}
m_{u_i \leftarrow v_{j}^k} = \gamma
(\textbf{h}_{v_j^{k}}, \rho_{i,j}^k);~~m_{v_{j}^k \leftarrow u_i} = \gamma(\textbf{h}_{u_i}, \rho_{i,j}^k)
\end{align}
\noindent where $\gamma(\cdot)$ denotes the information encoding function over the input feature embeddings $\textbf{h}_{v_j^{k}} \in \mathbb{R}^{(J\cdot K)\times d}$, $\textbf{h}_{u_i} \in \mathbb{R}^{I\times d}$. $\rho_{i,j}^k$ is the decay factor to normalize the propagated influence with node degrees~\cite{chen2020revisiting}, \ie, $\rho = \frac{1}{\sqrt{|N_i||N_{j}^k|}}$, where $N_i$ denotes the number of neighboring nodes of user $u_i$ and $N_{j}^k$ represents the number of connected user nodes of item $v_j$ under the relation type of $k$. Hence, the constructed message can be unfolded as:
\begin{align}
m_{u_i \leftarrow v_j^{k}} = \frac{1}{\sqrt{|N_i||N_{j}^k|}} (\textbf{h}_{v_j^{k}} \cdot \textbf{W}_1)
\end{align}
\noindent where $\textbf{W}_1 \in \mathbb{R}^{d\times d}$ is the weight matrix. We apply the similar operation for the message propagation from $u_i$ to type-specific item $v_j^{k}$.

\subsubsection{Temporal Context Encoding Scheme.}
Inspired by the recommendation techniques with the modeling of temporal information~\cite{sun2019bert4rec,huang2019online}, in our framework, we allow the user-item interactions happening at different timestamps interweave with each other, by introducing a temporal context encoding scheme to model the dynamic dependencies across different types of users' interactions. Motivated by the positional encoding algorithm in Transformer architecture~\cite{vaswani2017attention,sun2019bert4rec,wu2020hierarchically}, we map the timestamp $t_{i,j^k}$ of individual interaction $x_{i,j}^k$ into separate time slot as: $\widetilde{T}(t_{i,j}^k)$. We employ the sinusoid functions to generate the relative time embedding for edge $e_{i,j}^k \in E_m$ in $G_m$ as:
\begin{align}
\textbf{b}_{\widetilde{T}(t_{i,j}^k), 2i} = sin(\widetilde{T}(t_{i,j}^k)/10000^{\frac{2i}{d}}) \nonumber\\
\textbf{b}_{\widetilde{T}(t_{i,j}^k), 2i+1} = cos(\widetilde{T}(t_{i,j}^k)/10000^{\frac{2i+1}{d}}) 
\end{align}
\noindent where $(2i)$ and $(2i+1)$ denotes the element index with the even and odd position in embedding $\textbf{b}_{\widetilde{T}(t_{i,j}^k)}$, respectively.

\subsubsection{High-Order Message Aggregation Phase.}
We incorporate the propagated message between user $u_i$ and item $v_{i,j}^k$, as well as temporal context $\textbf{b}_{\widetilde{T}(t_{i,j}^k)}$ on their interaction edge $e_{i,j}^k$, into our information propagation paradigm as below:
\begin{small}
\begin{align}
\textbf{h}_{u_i}^{(l+1)} &= \varphi \Big (m^{(l)}_{u_i \leftarrow u_{i}} + \sum_{(j,k)\in N_{u_i}} m_{u_i \leftarrow v_{j}^k}^{(l)} \Big) = \varphi \Big ( \frac{1}{|N_{u_{i}}|} \textbf{h}_{u_{i}}^{(l)} \textbf{W}_2^{(l)} \nonumber\\
&+ \sum_{(j,k)\in N_{u_i}} \frac{1}{|N_{v_{j}^k}|} ( \textbf{h}_{v_{j}^k}^{(l)} \oplus \textbf{b}_{\widetilde{T}(t_{i,j}^k)} ) \textbf{W}_1^{(l)}) \Big ) 
 \end{align}
\end{small}

\noindent where $\varphi(\cdot)$ denotes the LeakyReLU function to perform the transformation. $m^{(l)}_{u_i \leftarrow u_{i}}$ is the self-propagated message with the weight matrix $\textbf{W}_2^{(l)} \in \mathbb{R}^{d\times d}$. $\oplus$ denotes the element-wise addition. $l$ is the index of $L$ graph layers.
We finally generate the user/item embeddings (\ie, $\textbf{h}_{u_i}^*$, $\textbf{h}_{v_{i,j}^k}^*$) with the following concatenation operation $\Vert$ as follows:
\begin{align}
\label{eq:highOrder_embed}
\textbf{h}_{u_i}^* &= (\textbf{h}_{u_i}^{(0)} \mathbin\Vert \textbf{h}_{u_i}^{(1)} \mathbin\Vert \cdots \mathbin\Vert \textbf{h}_{u_i}^{(L)})\nonumber\\
\textbf{h}_{v_{j,k}}^* &= (\textbf{h}_{v_{j,k}}^{(0)} \mathbin\Vert \textbf{h}_{v_{j,k}}^{(1)} \mathbin\Vert \cdots \mathbin\Vert \textbf{h}_{v_{j,k}}^{(L)})
\end{align}
\noindent We generate the summarized representation $\textbf{h}_{v_{j}}^*$ over all item sub-vertex embeddings $\textbf{h}_{v_{j,k}}^*$ ($k\in [1,...,K]$) with a gating mechanism~\cite{ma2019hierarchical}, to differentiate the importance of type-specific interaction patterns.

\subsection{Knowledge-aware Coupled Graph Neural Module}
\begin{figure}
	\centering
	\includegraphics[width=0.49\textwidth]{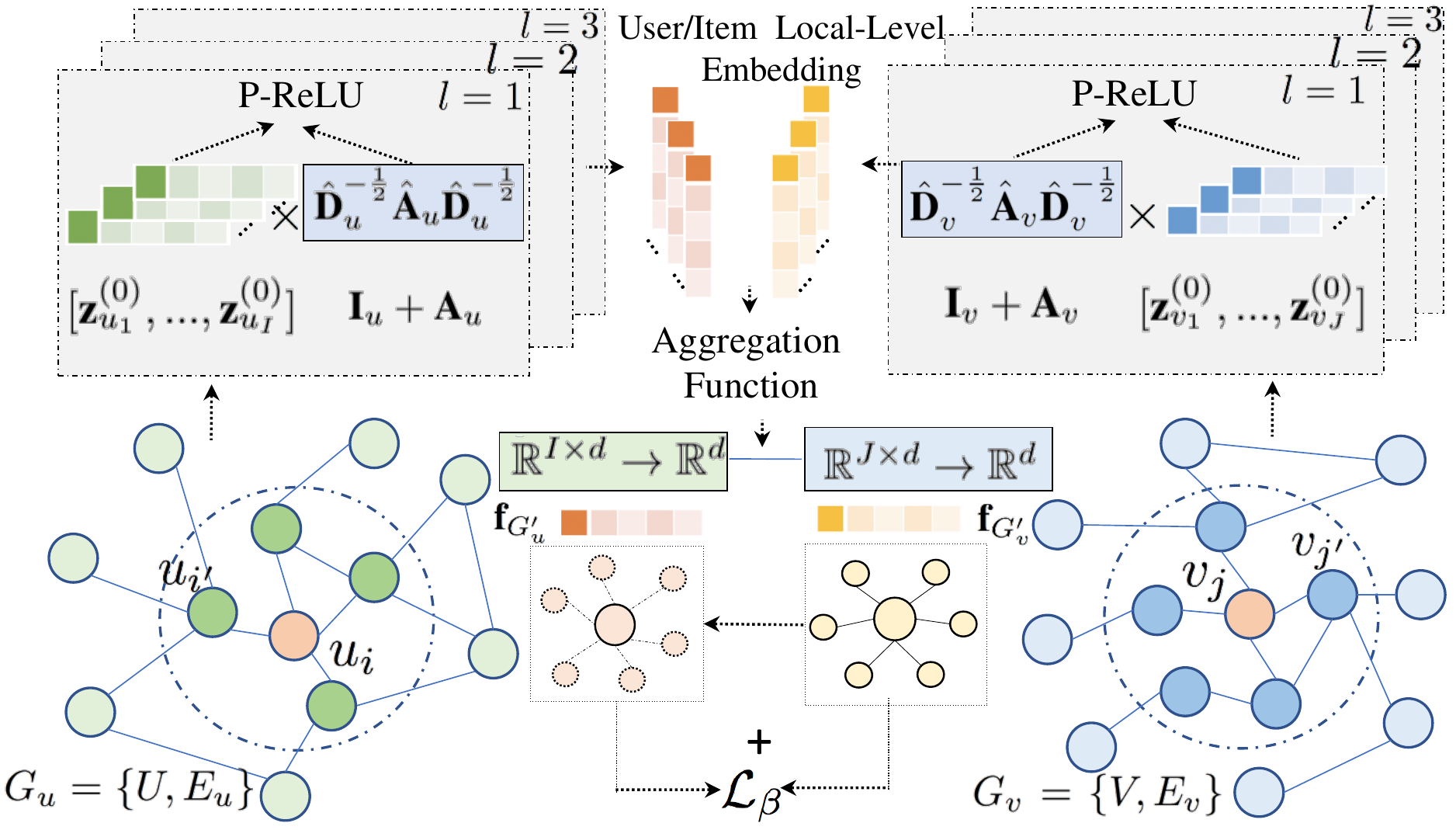}
	\caption{The architecture of joint encoding of user-user and item-item inter-dependent relational structures.}
	\label{fig:framework_2}
\end{figure}

To jointly inject the user- and item-wise inter-dependent knowledge into our user preference modeling, we develop a knowledge-aware coupled graph neural network which enables the collaboration between the mutual information learning and graph representation paradigm. While many efforts have been devoted to modeling graph structural information, they are limited in their ability in capturing both local and global graph substructure awareness~\cite{velickovic2019deep}, such as the user- and item-specific social/knowledge dependent information and high-order relationships across users/items. \model\ is equipped with a dual-stage graph learning paradigm (As shown in Figure~\ref{fig:framework_2}).

\subsubsection{Local Relational Structure Modeling.} We first learn the user- and item-specific specific embeddings ($\textbf{z}_{u_i}$, $\textbf{z}_{v_{j}}$) which preserves the local connection information over user social graph $G_u$ and item inter-dependent graph $G_v$ with the following graph-based update functions ($\textbf{z}_{u_i}^0$=$ \textbf{h}_{u_{i}}^*$, $\textbf{z}_{v_j}^0$ = $\textbf{h}_{v_{j}}^*$):
\begin{align}
[\textbf{z}_{u_1}^{(l+1)},...,\textbf{z}_{u_I}^{(l+1)}] =  \varphi \Big ([\textbf{z}_{u_1}^{(l)},...,\textbf{z}_{u_I}^{(l)}] \cdot \eta(G_u) \Big) \nonumber\\
[\textbf{z}_{v_1}^{(l+1)},...,\textbf{z}_{v_J}^{(l+1)}] = \varphi \Big ( [\textbf{z}_{v_1}^{(l)},...,\textbf{z}_{v_J}^{(l)}] \cdot \eta(G_v) \Big)
\end{align}
\noindent where $\eta(\cdot)$ denotes the adjacent relations of $G_u$ and $G_v$ with the symmetric normalization strategy in the information aggregation across the neighboring users/items, \eg, $\eta(G_v) = \hat{\textbf{D}}^{-\frac{1}{2}}_v \hat{\textbf{A}}_v \hat{\textbf{D}}^{-\frac{1}{2}}_v$. Hence, $\hat{\textbf{A}}_v$ is the addition of identity matrix $\textbf{I}_v$ and adjacent matrix $\textbf{A}_v$, so as to incorporate the information self-propagation~\cite{chen2020revisiting}.

In this graph learning paradigm, we aim to inject both local- and global-level relational structures over the user social graph and item relation graph into our learned user/item representations. Different from the existing graph neural network approaches~\cite{velickovic2019deep,icdm2020} which model the mutual relations between local feature embeddings and a single global representation, we enrich the global semantics with the consideration of connected graph substructures (\eg, the entire social relations of all users may consist of different connected sub-graphs $G'_u$). In particular, we first generate a fused graph-level representation $\textbf{f}_{G'_u}, \textbf{f}_{G'_v} \in \mathbb{R}^{d}$ by applying the mean pooling over node-specific embeddings.


We design our neural mutual information estimator based on a discriminator $\mathcal{D}(x,y)$ for node-graph pairwise relationships, to provide probability scores for sampled pairs. To be specific, we generate positive samples as $(\textbf{z}_{u_i}, \textbf{f}_{G'_u})$, $(\textbf{z}_{v_j}, \textbf{f}_{G'_v})$, and negative samples as $(\widetilde{\textbf{z}}_{u_i}, \textbf{f}_{G'_u})$, $(\widetilde{\textbf{z}}_{v_j}, \textbf{f}_{G'_v})$. Here, $\widetilde{\textbf{z}}_{u_i}$ and $\widetilde{\textbf{z}}_{u_i}$ are randomly picked with node shuffling to generate the misplaced node-graph pairwise relations.


Due to the rationality of cross-entropy in mutual information maximization~\cite{wang2020exploiting}, we define our noise-contrastive knowledge-aware loss function $\mathcal{L}_{\beta}$ as follows:
\begin{align}
\label{eq:dgi_loss}
\mathcal{L}_{\beta} &= - \frac{\lambda_1}{N_{pos}^u+N_{neg}^u} \Big ( \sum_{i=1}^{N_{pos}^u} \tau(\textbf{z}_{u_i}, \textbf{f}_{G'_u}) \cdot log \sigma(\textbf{z}_{u_i} \cdot \textbf{f}_{G'_u}) \nonumber\\
&+ \sum_{i=1}^{N_{neg}^u} \tau(\widetilde{\textbf{z}}_{u_i}, \textbf{f}_{G'_u}) \cdot log [1-\sigma(\widetilde{\textbf{z}}_{u_i} \cdot \textbf{f}_{G'_u})] \Big ) \nonumber\\
-& \frac{\lambda_2}{N_{pos}^v+N_{neg}^v} \Big ( \sum_{i=1}^{N_{pos}^v} \tau(\textbf{z}_{v_j}, \textbf{f}_{G'_v}) \cdot log \sigma(\textbf{z}_{v_j} \cdot \textbf{f}_{G'_v}) \nonumber\\
&+ \sum_{i=1}^{N_{neg}^v} \tau(\widetilde{\textbf{z}}_{v_j}, \textbf{f}_{G'_u}) \cdot log [1-\sigma(\widetilde{\textbf{z}}_{v_j} \cdot \textbf{f}_{G'_v})] \Big )
\end{align}
\noindent where $N_{pos}^u$/$N_{pos}^v$ and $N_{neg}^v$/$N_{neg}^v$ denotes the number of positive and negative instances sampled over sub-graph $G'_u$ and $G'_v$. $\tau(\cdot)$ is an indicator function, \eg, $\tau(\textbf{z}_{v_j}, \textbf{f}_{G'_v})=1$ and $\tau(\widetilde{\textbf{z}}_{v_j}, \textbf{f}_{G'_v})=1$ corresponds to the positive and negative pair instances. $\lambda_1$ and $\lambda_2$ are balance parameters. We aim to minimize $\mathcal{L}_{\beta}$ which is equivalent to maximize the mutual information, to jointly preserve the node-specific user/item characteristics and global graph-level dependencies.

\subsection{Model Optimization}
We define our loss $\mathcal{L}$ which includes (i) multi-typed user-item interaction encoding; (ii) knowledge-aware user-user and item-item inter-dependent relation learning. Particularly, $\mathcal{L}$ integrates the pairwise BPR loss, which is widely adopted in recommendation tasks~\cite{wang2019neural}, with the mutual information maximization paradigm as:
\begin{small}
\begin{align}
\mathcal{L} = \sum_{(i,j^+,j^-)\in O} -\text{In}~\sigma(\widehat{x}_{i,j^+} - \widehat{x}_{i,j^-}) + \lambda \mathbin\Vert \Theta \mathbin\Vert^2  + \mathcal{L}_{\beta}
\end{align}
\end{small}
\noindent the pairwise training data is denoted as $O=\{(u,j^+,j^-) | (u, j^+) \in \mathcal{R}^+, (u, j^-) \in \mathcal{R}^-\}$ ($\mathcal{R}^+$, $\mathcal{R}^-$ denotes the observed and unobserved interactions, respectively). $\widehat{x}_{i,j}$ is the calculated score with the inner-product between the embedding of $u_i$ and $v_j$. $\Theta$ are trainable parameters, $\sigma(\cdot)$--sigmoid. $\lambda$ controls the strength of $L_2$ regularization for overfitting alleviation.

\subsubsection{Time Complexity Analysis.} \model\ takes $O(|E|\times d)$ for the message passing in handling the user-user, user-item and item-item relations, where $|E|$ denotes the number of edges. Also, $O((I+J\cdot K)\times d^2)$ computation is spent by the transformations. Typically, the first term is dominant due to information compression. In conclusion, \model\ is comparable in time efficiency compared with current GNN-based recommendation methods. Our model only utilizes moderate memory to store node embeddings ($O((I+J\cdot K)\times d)$), which is similar to the existing methods.

\section{Evaluation}
\label{sec:eval}

In this section, we conduct experiments on different real-world datasets to evaluate the performance of our method from the following aspects:
\begin{itemize}
\item \textbf{RQ1}: Does \emph{\model} consistently outperform other baseline in terms of recommendation accuracy?
\item \textbf{RQ2}: How is the performance of \emph{\model}'s variants with the combination of different relation encoders?
\item \textbf{RQ3}: How is forecasting performance of compared methods \wrt\ different interaction density degrees? 
\item \textbf{RQ4}: How do the representations benefit from the collectively encoding of global knowledge-aware cross-interactive patterns in social recommendation?
\item \textbf{RQ5}: How do different hyper-parameter settings impact the performance of our \emph{\model} framework?
\item \textbf{RQ6}: How is the model efficiency of the \emph{\model}?
\end{itemize}

\subsection{Experimental Settings}
\subsubsection{\bf Dataset.} Table~\ref{tab:data} lists the statistics of three datasets. We describe the details of those datasets as follows:

\noindent \textbf{Epinions}. This data records the user's feedback over different items from a social network-based review system Epinions~\cite{fan2019graph}. Each explicit rating score (ranging from 1 to 5) is regarded as an individual type of interaction: negative, below average, neutral, above average, positive.

\noindent \textbf{Yelp}. This data is collected from the Yelp platform, in which user-item interactions are differentiated with the same split rubric in Epinions. Furthermore, user's social connections (with common interests) are contained in this data.

\noindent \textbf{E-Commerce}. It is collected from a commercial e-commerce platform with different types of interactions, \ie, \emph{page view}, \emph{add-to-cart}, \emph{add-to-favorite} and \emph{purchase}. User's relations are constructed with their co-interact patterns.

The item inter-dependency graph $G_v$ on the above datasets are constructed based on item categories.


\begin{table}[t]
\centering
\small
\begin{tabular}{l| c| c| c}
\hline
Dataset & Epinions & Yelp & E-commerce\\
\hline
\# of Users & 18,081 & 43,043 & 334,042 \\
\# of Items & 251,722 & 66,576 & 195,940\\
\# of Interactions & 715,821 & 283,512 & 1,930,466\\
\hline
\end{tabular}
\caption{Statistics of Experimented Datasets.}
\label{tab:data}
\end{table}


\begin{table*}[t]
\small
\begin{center}
\setlength{\tabcolsep}{1.0mm}
\begin{tabular}{| c || c| c | c | c | c | c | c | c | c | c | c | c || c|}
\hline
Data & Metrics & ~PMF~ & TrustMF & DiffNet & SAMN & DGRec & EATNN & NGCF+S & KGAT & GraphRec & Danser & LR-GCCF & \emph{\model} \\
\hline
\multirow{2}{*}{Epinions} & HR &0.619 & 0.635 & 0.632 & 0.639 & 0.626 & 0.642 & 0.707 & 0.675 & 0.686 & 0.669  &0.677 &\textbf{0.742}	\\
\cline{2-14}
& NDCG  &  0.410 & 0.417 & 0.416 & 0.425 & 0.412 & 0.448 & 0.498 & 0.470 & 0.478 & 0.462 &0.478 &\textbf{0.513} \\
\cline{2-14}
\hline
\multirow{2}{*}{Yelp} & HR &0.698 & 0.756 & 0.785 & 0.751 & 0.766 & 0.771 & 0.781 & 0.772 & 0.7605 & 0.774 & 0.769& 
 \textbf{0.8026}\\
\cline{2-14} 
& NDCG &0.460 & 0.495 & 0.512 & 0.486 & 0.495 & 0.506 & 0.523 & 0.511 & 0.494 & 0.508 &0.518 & 
\textbf{0.530} \\
\cline{1-14}
\multirow{2}{*}{E-Com} & HR &0.654 & 0.674 & 0.722 & 0.676 & 0.672 & 0.683 & 0.694 & 0.689 & 0.668 & 0.670& 0.690& \textbf{0.735}\\
\cline{2-14}
& NDCG  &0.431 & 0.452 & 0.519 & 0.461 & 0.441 & 0.456 & 0.476 & 0.473 & 0.439 & 0.443& 0.485& \textbf{0.529}\\
\cline{1-14}
\hline
\end{tabular}
\end{center}
\vspace{-0.15in}
\caption{Performance comparison of all methods in interaction prediction.} 
\label{tab:result}
\end{table*}


\subsubsection{\bf Evaluation Protocols.}
We adopt two widely used evaluation metrics for social recommendation tasks~\cite{chen2019social}: \textit{Hit Ratio (HR@$N$)} and \textit{Normalized Discounted Cumulative Gain (NDCG@$N$)}. We follow the evaluation settings in~\cite{chen2019efficient,wu2019neural} and employ the leave-one-out method for generating training and test data instances. To be consistent with~\cite{sun2019bert4rec}, we associate each positive instance with 99 negative samples.

\subsubsection{\bf Baselines.}
In our experiments, we perform the performance comparison by considering the following baselines:

\noindent \textbf{Probabilistic Matrix Factorization Method}.
\begin{itemize}[leftmargin=*]
\item \textbf{PMF}~\cite{mnih2008probabilistic}: it is a probabilistic approach with the matrix factorization for user/item factorization.
\end{itemize}

\noindent \textbf{Conventional Social Recommendation Methods}.
\begin{itemize}[leftmargin=*]
\item \textbf{TrustMF}~\cite{yang2016social}: this method incorporates the truth relationships between users into the matrix factorization architecture for user interaction embedding.
\end{itemize}

\noindent \textbf{Attentive Social Recommendation Techniques}.
\begin{itemize}[leftmargin=*]
\item \textbf{SAMN}~\cite{chen2019social}: this model is a dual-stage attention network which learns the influences between the target user and his/her neighboring nodes.
\item \textbf{EATNN}~\cite{chen2019efficient}: This transfer learning model is also on the basis of attention mechanism to jointly fuse information from user's interactions and social signals.
\end{itemize}

\noindent \textbf{Graph Neural Networks Social Recommender Systems}.
\begin{itemize}[leftmargin=*]
\item \textbf{DiffNet}~\cite{wu2019neural}: it is a deep influence propagation framework to model the social diffusion process.\vspace{-0.05in}
\item \textbf{GraphRec}~\cite{fan2019graph}: it aggregates the social relations between users via a graph neural architecture.\vspace{-0.05in}
\item \textbf{NGCF+S}~\cite{wang2019neural}: we incorporate the social ties into the state-of-the-art graph-structured neural collaborative filtering model for joint message propagation.\vspace{-0.05in}
\item \textbf{Danser}~\cite{wu2019dual}: it is composed of two graph attention layers for capturing the social influence and homophily, respectively from both users and items.\vspace{-0.05in}
\item \textbf{LR-GCCF}~\cite{chen2020revisiting}: it is a new graph-based collaborative filtering model based on graph convolutional network by removing non-linear transformations.
\end{itemize}

\noindent \textbf{Social Recommendation with Sequential Pattern}.
\begin{itemize}[leftmargin=*]
\item \textbf{DGRec}~\cite{song2019session}: it jointly models the dynamic user's preference and the underlying social relations.
\end{itemize}

\noindent \textbf{Knowledge Graph-enhanced Recommendation}.
\begin{itemize}[leftmargin=*]
\item \textbf{KGAT}~\cite{wang2019kgat}: it is a graph attentive message passing framework which utilizes the knowledge graph to enhance the recommendation with item side information.
\end{itemize}

\subsubsection{\bf Implementation Details.}
The \emph{\model} is implemented with Pytorch and Adam optimizer is adopted for hyperparameter estimation. The training process is performed with the learning rate range of [0.001, 0.005, 0.01], and the batch size selected from $[1024, 2048, 4096, 8192]$. The embedding size is tuned from the range of $[8,16,32,64]$. In our evaluations, we employ the early stopping for training termination when the performance degrades for 5 continuous epochs on the validation data.

\begin{table}[t]
\small
\begin{center}
\setlength{\tabcolsep}{0.4mm}
\begin{tabular}{| c | c | c | c | c | c | c | c|}
\hline
Data & Metrics & DiffNet & DGRec & KGAT & GraphRec & Danser & \emph{\model} \\
\hline
\multirow{2}{*}{Epinions} & HR & 0.628 &0.625 & 0.685 & 0.678 & 0.653  &\textbf{0.745}	\\
\cline{2-8}
& NDCG  & 0.411 & 0.409 & 0.480 & 0.465 & 0.444 &\textbf{0.519} \\
\cline{2-8}
\hline 
\multirow{2}{*}{Yelp} & HR & 0.809 & 0.808 & 0.791 & 0.781 & 0.790 &\textbf{0.839}\\
\cline{2-8} 
& NDCG & 0.542 & 0.534 & 0.530 & 0.520 & 0.533 &\textbf{0.573} \\
\cline{1-8}
\multirow{2}{*}{E-Com} & HR & 0.894 & 0.900 & 0.886 & 0.849 & 0.872 & \textbf{0.911}\\
\cline{2-8}
& NDCG & 0.673 & 0.659 & 0.653 & 0.627 & 0.649 & \textbf{0.710}\\
\cline{1-8}
\hline
\end{tabular}
\end{center}
\caption{Prediction results for like/purchase activities on three datasets in terms of  \emph{HR@10} and \emph{NDCG@10}.}
\label{tab:result_target}
\end{table}


\subsection{Overall Model Performance Comparison (RQ1)}
Table~\ref{tab:result} reports the results of \emph{\model} and many baselines in predicting the overall interactions in terms of HR@10 and NDCG@10. It can be seen that \emph{\model} consistently obtains the best performance across different recommendation scenarios in terms of two metrics, which justifies the effectiveness of our method in integrating user-user and item-item relations, with the multi-typed user-item interactive patterns.

Compared with traditional approaches, neural network based models usually achieve better performance, due to the modeling of high-level non-linearities during the feature interaction learning phase. Among various compared approaches, the GNN-based models outperforms the attentive social recommender systems, which ascertains the rationality of applying graph neural networks for high-order relations across users/items in a recursive way. Different from those GNN techniques, our framework integrates the social and knowledge-aware relations from global context via a mutual information encoding paradigm, and also captures interaction dynamics, which results in better performance.\\\vspace{-0.12in}

We further investigate the performance of our \emph{\model} in making recommendations on the target type of interactions (\eg, positive feedback on Epinions and Yelp or user's purchase on E-commerce). The results are shown in Table~\ref{tab:result_target}. We can observe that \emph{\model} still achieves significant improvement, with the careful consideration of different types of user-item interaction signals. While the baseline KGAT proposes to incorporate the auxiliary knowledge graph, it fails to explicitly differentiate type-specific interaction patterns.\\\vspace{-0.12in}

\subsection{Impact of Different Relation Encoders (RQ2)}
We next perform experiments to evaluate the impact of the incorporation of multi-typed user-item interactions, user-wise relations, item-wise dependencies, and the temporal context, with the following five contrast variants of \emph{\model}.
\begin{itemize}[leftmargin=*]
\item \emph{\model}-M: \emph{\model} without modeling multi-typed interaction patterns and only with singular-type interactions. \vspace{-0.05in}
\item \emph{\model}-U: \emph{\model} without the social relation encoder for capturing the social signals in the recommendation. \vspace{-0.05in}
\item \emph{\model}-I: \emph{\model} without the external knowledge to characterize the item dependency. \vspace{-0.05in}
\item \emph{\model}-UI: \emph{\model} without both the user- and item-wise relation encoders and remove the coupled mutual information paradigms in the joint learning framework.\vspace{-0.05in}
\item \emph{\model}-T: \emph{\model} without the temporal context encoding.
\end{itemize}

\begin{figure}[t]
	\centering
	\subfigure[][Epinions]{
		\centering
		\includegraphics[width=0.29\columnwidth]{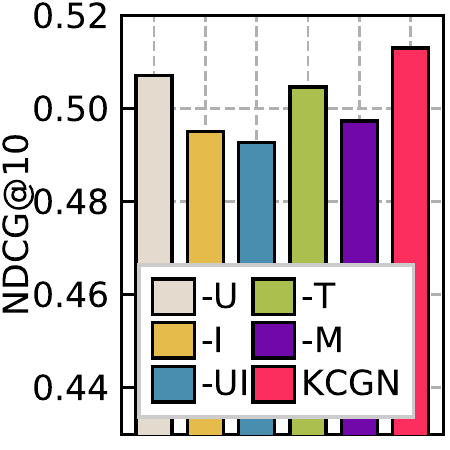}
		\label{fig:ab_yelp_NDCG}
	}
	\subfigure[][Yelp]{
		\centering
		\includegraphics[width=0.29\columnwidth]{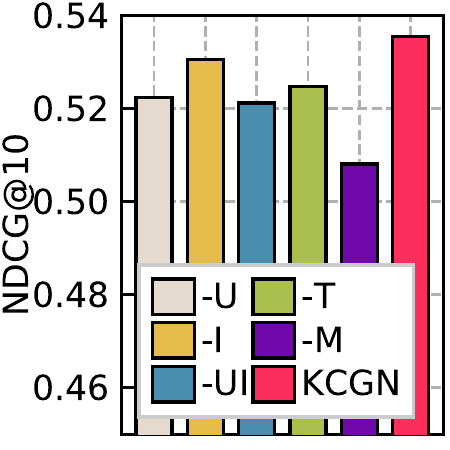}
		\label{fig:ab_ml10m_NDCG}
	}
	\subfigure[][E-commerce]{
		\centering
		\includegraphics[width=0.29\columnwidth]{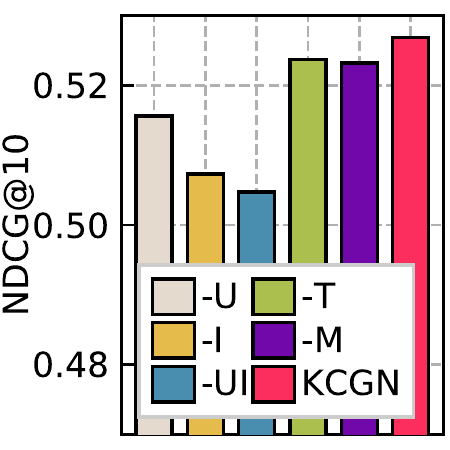}
		\label{fig:ab_retail_NDCG}
	}
	\caption{Ablation studies for different sub-modules of \model\ framework, in terms of \textit{HR@10} and \textit{NDCG@10}.}
	\label{fig:ablation}
\end{figure}

Figure~\ref{fig:ablation} shows the comparison results of different variants. We can see that the joint model \emph{\model} achieves the best performance. As such, it is necessary to build a joint framework to simultaneously capture social dimension (users' social influence), item dimension (knowledge-aware inter-item relations), multi-typed interactions, and time-aware user's interest, for making recommendations. In addition, \emph{\model}-UI performs worse than \emph{\model}-U and \emph{\model}-I, which again confirms the efficacy of our designed relation aggregation functions.

\subsection{Performance over Sparsity Degrees (RQ3)}
\begin{figure}[t]
	\centering
	\subfigure[][Yelp]{
		\centering
		\includegraphics[width=0.47\columnwidth]{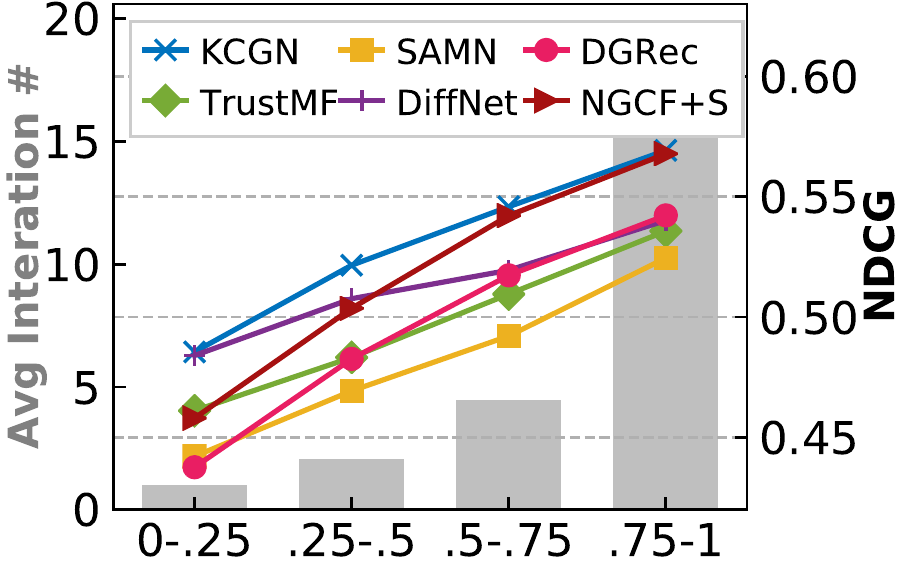}
		\label{fig:embed_NGCF}
	}
	\subfigure[][E-Commerce]{
		\centering
		\includegraphics[width=0.47\columnwidth]{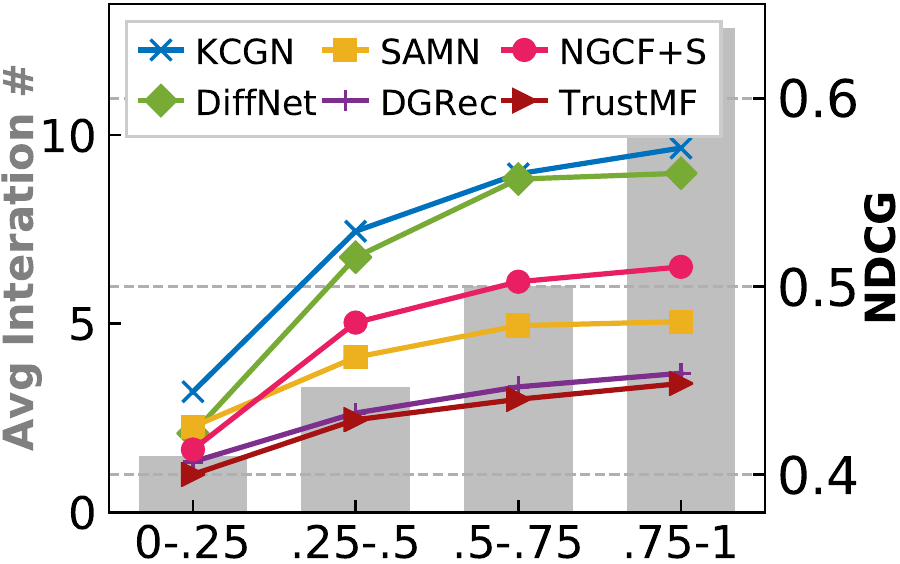}
		\label{fig:embed_KCGN}
	}
	\vspace{-0.05in}
	\caption{Performance of \model\ and baselines over users with different sparsity from Yelp and E-Commerce data.}
	\label{fig:sparsity}
\end{figure}

One key motivation to exploit social- and knowledge-aware side information is to alleviate the sparsity issue, which limits the model robustness. Hence, we further evaluate our \emph{\model} for both inactive and active users. In particular, we partition the target users into four sparsity levels in terms of their interaction densities. Figure~\ref{fig:sparsity} presents the evaluation results on different user groups on Yelp and E-Commerce data in terms of \textit{NDCG@10}. We can observe that \emph{\model} outperforms representative baselines in most cases, especially on sparest user groups. This suggests that incorporating both user and item side knowledge as their external relations, empowers the representations of inactive users through our recursive information aggregation architecture.

\subsection{Qualitative Analyses of \emph{\model} (RQ4)}
\begin{figure}[t]
	\centering
	\subfigure[][NGCF+S]{
		\centering
		\includegraphics[width=0.45\columnwidth]{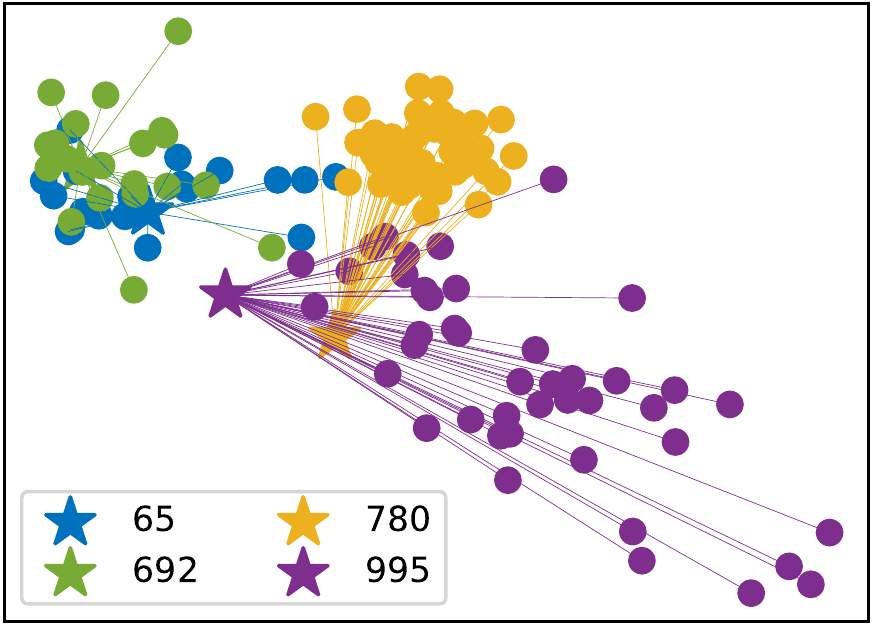}
		\label{fig:embed_NGCF}
	}
	\subfigure[][\model]{
		\centering
		\includegraphics[width=0.45\columnwidth]{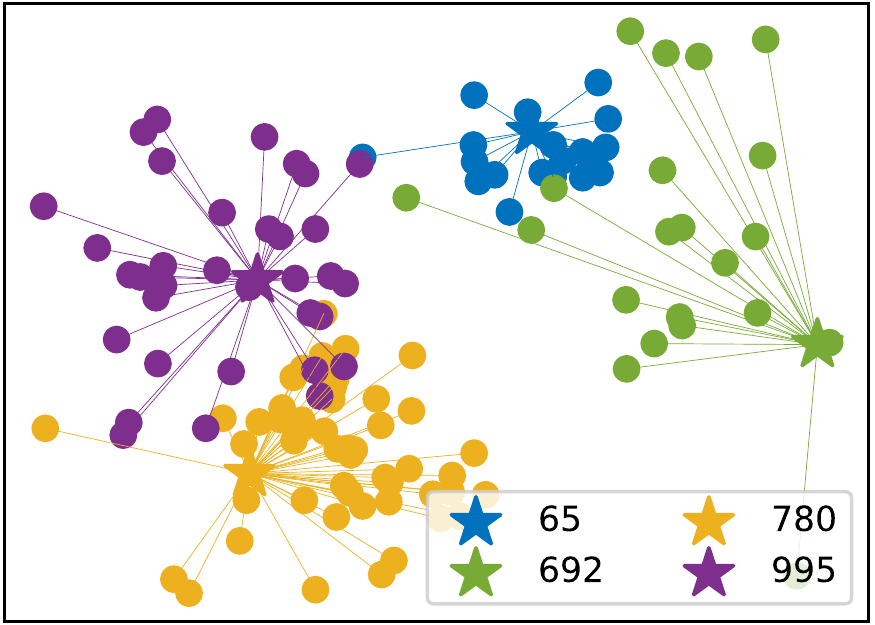}
		\label{fig:embed_KCGN}
	}
	\caption{Visualized embeddings for users (stars) and their 4- or 5-rated item (circles), learned by \model\ and NGCF+S.}
	\label{fig:embed_case}
\end{figure}

We illustrate how our side knowledge-aware multi-typed relation encoding schema benefit the ability of embedding user's preference into the latent learning space. In particular, we sample several users and their four- and five-star rated items from Yelp dataset, and further visualize the corresponding user/item embeddings learned by NGCF+S and our \model\ (as shown in Figure~\ref{fig:embed_case}). From the results, we can notice that: i) the visualized embeddings could well preserve the relationships between users and their interacted items with a clustering phenomenon (represented with the same color); ii) \emph{\model} could provide a better separation for different users and their interacted items. Hence, the above observations verify the superior representation learning ability of \model\ through the encoding function which maps the side knowledge and interaction units into effective latent space.

\subsection{Parameter Sensitivity Study (RQ5)}
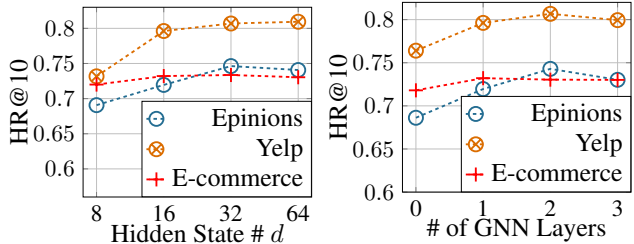
\begin{figure}[t]
    \centering
    \begin{adjustbox}{max width=1.0\linewidth}

\begin{tikzpicture}
\begin{axis}[
    xlabel={Hidden State \# $d$},
    ylabel={HR@10},
    xmin=0.8,xmax=4.2,
    ymin=0.56,ymax=0.83,
    xtick={1,2,3,4},
    xticklabels={8,16,32,64},
    legend columns=1,
    legend cell align=right,
    grid=both,
    every axis plot/.append style={ultra thick},
    every tick label/.append style={scale=2},
    label style={scale=2.2},
    ylabel style={yshift=2ex},
    legend style={
        nodes={scale=2, transform shape},
        },
    legend style={at={(1,0)},anchor=south east},
    every axis plot post/.append style={
        every mark/.append style={scale=2}
    }
]

\addplot[color={rgb:blue,4;green,2;yellow,1}, mark=o, dashed, mark options={solid, scale=1.5}]
table[x=para, y=ep_hr] {latFactor.txt};
\addplot[color={rgb:red,4;green,1;yellow,2}, mark=otimes, dashed, mark options={solid, scale=1.5}]
table[x=para, y=yelp_hr] {latFactor.txt};
\addplot[color={rgb:red,4}, mark=+, dashed, mark options={solid, scale=1.5}]
table[x=para, y=ecom_hr] {latFactor.txt};
\legend{\large Epinions, \large Yelp, \large E-commerce};
\end{axis}
\end{tikzpicture}

\begin{tikzpicture}
\begin{axis}[
    xlabel={\# of GNN Layers},
    ylabel={HR@10},
    xtick={1,2,3,4},
    xticklabels={0,1,2,3},
    xmin=0.8,xmax=4.2,
    ymin=0.6,ymax=0.82,
    legend columns=1,
    legend cell align=right,
    grid=both,
    every axis plot/.append style={ultra thick},
    every tick label/.append style={scale=2},
    label style={scale=2.2},
    ylabel style={yshift=2ex},
    legend style={
        nodes={scale=2, transform shape},
        },
    legend style={at={(1,0)},anchor=south east},
    every axis plot post/.append style={
        every mark/.append style={scale=2}
    }
]

\addplot[color={rgb:blue,4;green,2;yellow,1}, mark=o, dashed, mark options={solid, scale=1.5}]
table[x=para, y=ep_hr] {gnn_layer.txt};
\addplot[color={rgb:red,4;green,1;yellow,2}, mark=otimes, dashed, mark options={solid, scale=1.5}]
table[x=para, y=yelp_hr] {gnn_layer.txt};
\addplot[color={rgb:red,4}, mark=+, dashed, mark options={solid, scale=1.5}]
table[x=para, y=ecom_hr] {gnn_layer.txt};
\legend{\large Epinions, \large Yelp, \large E-commerce};

\end{axis}
\end{tikzpicture}



    \end{adjustbox}
    \caption{Hyper-parameter study of \emph{\model}}
    \label{fig:hyperparam}
\end{figure}

\noindent \textbf{Impact of \# Recursive Graph Layers}.
Figure~\ref{fig:hyperparam} shows the experimental results with different number of embedding propagation layers over user-item interaction graph. We can observe that increasing the depth of \emph{\model} could boost the performance, \ie, \emph{\model}-2 performs better than \emph{\model}-0 (without the graph structure) and \emph{\model}-1 (only consider 1-hop neighbors). The performance improvement lies in the effective modeling of high-order collaborative effects across users and items. \emph{\model} with 3 graph layers performs worse than \emph{\model}-2, suggests that exploring higher-level relations may involve noise.

\noindent \textbf{Impact of Embedding Dimension}. We notice that the accuracy is initially improved with larger embedding size due to the stronger representation ability. However, the performance degrades with the further increase of dimensionality, which indicates the overfitting phenomenon.


\subsection{Model Efficiency Study (RQ6)}
We finally investigate the computation cost of our \emph{\model} when competing with state-of-the-art baselines. As shown in Table~\ref{tab:time}, we can observe that \emph{\model} achieves competitive time efficiency (measured by running time of each epoch) when compared with neural social recommendation methods. It is worthwhile pointing out that methods with stacking multiple graph attention layers is time-consuming, due to their pairwise attentive weights calculations for social or knowledge graph information aggregation. 




\begin{table}
	\small
	\label{tab:run_time}
	\centering
    \setlength{\tabcolsep}{1mm}
	\begin{tabular}{|c|c|c|c|c|c|c|}
		\hline
		Data  & DGRec & SAMN & EATNN & KGAT &  GraphRec & \emph{\model}\\
		\hline
	    Epinions & 4.4 & 4.7 & 10.7 & 60.5 & 40.5 & 17.5\\
        \hline
		Yelp & 2.6 & 8.9 & 13.5 & 20.9& 7.3& 3.7\\
		\hline
		E-Com & 82.5 & 78.3 & 152.7 & 342.8& 497.3& 70.2\\
		\hline
	\end{tabular}
	\caption{Model computational cost with running time (s).}
	\label{tab:time}
\end{table}


\section{Related Work}
\label{sec:relate}

\noindent \textbf{Social-aware Recommender Systems}.
Deep learning has been revolutionizing recommender systems and many neural network models have been proposed for social recommendation scenario~\cite{yin2019social,chen2020social}. For example, attention mechanisms are introduced to learn the influences between users, such as SAMN~\cite{chen2019social} and EATNN~\cite{chen2019efficient}. It is worth mentioning that several recent efforts explore the GNNs for incorporating social relations into the user-item interaction encoding~\cite{wu2019dual,fan2019graph,wu2019neural,icdm2020}. Different from these methods, \model\ focus on fusing the heterogeneous relations from different aspects (social, item knowledge and temporal), to boost the performance.\\\vspace{-0.1in}

\noindent \textbf{Graph Methods for Recommendation}. Many recent efforts have been devoted to exploring insights from GNNs for modeling collaborative signals in recommender systems. For example, inspired by the graph convolutional operations, PinSage~\cite{ying2018graph} and NGCF~\cite{wang2019neural} aim to aggregate high-hop neighboring feature information over the user-item interaction graph. Several subsequent extensions have been developed to revisit the graph-based CF effects, such as LightGCN~\cite{he2020lightgcn}, LR-GCCF~\cite{chen2020revisiting} and KHGT~\cite{multibehavior2021graph}. Motivated by these works, we propose a new knowledge-aware graph neural architecture for social recommendation.

\section{Conclusion}
\label{sec:conclusion}

In this paper, we propose \model, an end-to-end framework that naturally incorporates knowledge-aware item dependency into the social recommender systems. \model\ unifies the user-user and item-item relation structure learning with a coupled graph neural network under a mutual information-based neural estimator. To handle the dynamic user-item interaction heterogeneity, we design a relation-aware graph encoder to empower \model\ to maintain dedicated representations of multi-typed interaction signals with the incorporation of temporal information. Through extensive experiments on real-world datasets, we demonstrate that \model\ achieves substantial gains over state-of-the-art baselines.

\section*{Acknowledgments}
We thank the anonymous reviewers for their constructive feedback and comments. This work is supported by National Nature Science Foundation of China (62072188, 61672241), Natural Science Foundation of Guangdong Province (2016A030308013), Science and Technology Program of Guangdong Province (2019A050510010). 

\bibliography{refs}

\end{document}